\documentclass[12pt,preprint]{aastex}

\usepackage[dvipdfm,bookmarks=true,bookmarksnumbered=true,bookmarkstype=toc,
colorlinks=true,linkcolor=black,urlcolor=blue,citecolor=blue]{hyperref}

\begin{document}

\title{A possible binary system of a stellar remnant \\
in the high magnification gravitational microlensing event \\
OGLE-2007-BLG-514}

\author{ 
N.~Miyake\altaffilmark{1}, 
A.~Udalski\altaffilmark{2}, 
T.~Sumi\altaffilmark{3}, 
D.~P.~Bennett\altaffilmark{4}, 
S.~Dong\altaffilmark{5}, 
R.~A.~Street\altaffilmark{6}, 
J.~Greenhill\altaffilmark{7}, 
I.~A.~Bond\altaffilmark{8}, 
A.~Gould\altaffilmark{5}, 
%%%%%%
%
\\and\\
%
%\if0
%%%%%%
% OGLE
M.~Kubiak\altaffilmark{2}, 
M.~K.~Szyma\'nski\altaffilmark{2}, 
G.~Pietrzy\'nski\altaffilmark{2,9}, 
I.~Soszy\'nski\altaffilmark{2},
K.~Ulaczyk\altaffilmark{2},  
L.~Wyrzykowski\altaffilmark{2,10} \\
(The OGLE Collaboration) \\
% MOA
F.~Abe\altaffilmark{1}, 
A.~Fukui\altaffilmark{11}, 
K.~Furusawa\altaffilmark{1}, 
S.~Holderness\altaffilmark{12}, 
Y.~Itow\altaffilmark{1}, 
A.~Korpela\altaffilmark{13}, 
C.~H. Ling\altaffilmark{8}, 
K.~Masuda\altaffilmark{1}, 
Y.~Matsubara\altaffilmark{1}, 
Y.~Muraki\altaffilmark{1}, 
T.~Nagayama\altaffilmark{14}, 
K.~Ohnishi\altaffilmark{15}, 
N.~Rattenbury\altaffilmark{16}, 
To.~Saito\altaffilmark{17}, 
T.~Sako\altaffilmark{1}, 
D.~J.~Sullivan\altaffilmark{18}, 
W.~L.~Sweatman\altaffilmark{8}, 
P.~J.~Tristram\altaffilmark{13}, 
P.~C.~M.~Yock\altaffilmark{16} \\
(The MOA Collaboration) \\
%
% uFUN
W.~Allen\altaffilmark{19}, 
G.~W.~Christie\altaffilmark{20},
D.~L.~DePoy\altaffilmark{5}, 
B.~S.~Gaudi\altaffilmark{5}, 
C.~Han\altaffilmark{21}, 
C.-U.~Lee\altaffilmark{22}, 
J.~McCormick\altaffilmark{23}, 
B.~Monard\altaffilmark{19}, 
T.~Natusch\altaffilmark{24}, 
B.-G.~Park\altaffilmark{22}, 
R.~W.~Pogge\altaffilmark{5} \\
(The $\mu$FUN Collaboration) \\
%
% RoboNet
% 2007 RoboNet team:
A.~Allan\altaffilmark{25}, 
M.~Bode\altaffilmark{26}, 
D.~M.~Bramich\altaffilmark{27}, 
N.~Clay\altaffilmark{26}, 
M.~Dominik\altaffilmark{28}, 
K.~D.~Horne\altaffilmark{28}, 
N.~Kains\altaffilmark{27}, 
C.~Mottram\altaffilmark{26}, 
C.~Snodgrass\altaffilmark{29,30}, 
I.~Steele\altaffilmark{26}, 
Y.~Tsapras\altaffilmark{6,31} \\
(The RoboNet Collaboration) \\
%
% PLANET
M.~D.~Albrow\altaffilmark{32}, 
V.~Batista\altaffilmark{33}, 
J.~P.~Beaulieu\altaffilmark{33}, 
S.~Brillant\altaffilmark{34}, 
M.~Burgdorf\altaffilmark{24,35}, 
J.~A.~R.~Caldwell\altaffilmark{36},
A.~Cassan\altaffilmark{33}, 
A.~Cole\altaffilmark{7}, 
K.~H.~Cook\altaffilmark{37}, 
Ch.~Coutures\altaffilmark{38}, 
S.~Dieters\altaffilmark{7,33}, 
D.~Dominis~Prester\altaffilmark{39}, 
J.~Donatowicz\altaffilmark{40},
P.~Fouqu\'e\altaffilmark{41}, 
U.~G.~Jorgensen\altaffilmark{42}, 
S.~Kane\altaffilmark{43}, 
D.~Kubas\altaffilmark{40}, 
J.~B.~Marquette\altaffilmark{33},
R.~Martin\altaffilmark{44}, 
J.~Menzies\altaffilmark{45}, 
K.~R.~Pollard\altaffilmark{32}, 
K.~C.~Sahu\altaffilmark{46}, 
J.~Wambsganss\altaffilmark{47}, 
A.~Williams\altaffilmark{44}, 
M.~Zub\altaffilmark{47} \\
(The PLANET Collaboration) \\
}

\altaffiltext{1}{ Solar-Terrestrial Environment Laboratory, Nagoya University, Nagoya, 464-8601, Japan; nmiyake@stelab.nagoya-u.ac.jp}
\altaffiltext{2}{ Warsaw University Observatory, Al. Ujazdowskie 4, 00-478 Warszawa, Poland }
\altaffiltext{3}{ Department of Earth and Space Science, Osaka University, Osaka 560-0043, Japan }
\altaffiltext{4}{ University of Notre Dame, Department of Physics, 225 Nieuwland Science Hall, Notre Dame, IN 46556, USA}
\altaffiltext{5}{ Department of Astronomy, Ohio State University, 140 W. 18th Ave., Columbus, OH 43210, USA}
\altaffiltext{6}{ Las Cumbres Observatory, 6740B Cortona Dr, Suite 102, Goleta, CA 93117, USA }
\altaffiltext{7}{ University of Tasmania, School of Maths and Physics, Private bag 37, GPO Hobart, Tasmania 7001, Australia}
\altaffiltext{8}{ Institute for Information and Mathematical Sciences, Massey University, Private Bag 102-904, Auckland 1330, New Zealand}
\altaffiltext{9}{ Universidad de Concepcion, Departamento de Fisica, Casilla 160.C, Concepcion, Chile }
\altaffiltext{10}{ Institute of Astronomy, University of Cambridge, Madingley Road, Cambridge CB3 0HA, UK} 
\altaffiltext{11}{ Okayama Astrophysical Observatory, National Astronomical Observatory of Japan, Okayama 719-0232, Japan }
\altaffiltext{12}{ Computer Science Department, University of Auckland, Auckland, New Zealand}
\altaffiltext{13}{ Mt. John Observatory, PO Box 56, Lake Tekapo 8770, New Zealand}
\altaffiltext{14}{ Department of Physics and Astrophysics, Faculty of Science, Nagoya University, Nagoya 464-8602, Japan}
\altaffiltext{15}{ Nagano National College of Technology, Nagano 381-8550, Japan}
\altaffiltext{16}{ Department of Physics, University of Auckland, Private Bag 92-019, Auckland 1001, New Zealand }
\altaffiltext{17}{ Tokyo Metropolitan College of Industrial Technology, Tokyo 116-8523, Japan}
\altaffiltext{18}{ School of Chemical and Physical Sciences, Victoria University, Wellington, New Zealand}
\altaffiltext{19}{ Bronberg Observatory, Centre for Backyard Astrophysics Pretoria, South Africa }
\altaffiltext{20}{ Auckland Observatory, Auckland, New Zealand }
\altaffiltext{21}{ Program of Brain Korea, Department of Physics, Chungbuk National University, 410 Seongbong-Rho, Hungduk-Gu, Chongju 371-763, Korea }
\altaffiltext{22}{ Korea Astronomy and Space Science Institute, 61-1 Hwaam-Dong, Yuseong-Gu, Daejeon 305-348, Korea }
\altaffiltext{23}{ Farm Cove Observatory, Centre for Backyard Astrophysics, Pakuranga, Auckland New Zealand }
\altaffiltext{24}{ AUT University, Auckland, New Zealand }
\altaffiltext{25}{ School of Physics, University of Exeter, Stocker Road, Exeter, Devon, EX4 4QL, England, UK }
\altaffiltext{26}{ Astrophysics Research institute, Liverpool John Moores University, Egerton Wharf, Birkenhead CH41 1LD, England, UK }
\altaffiltext{27}{ European Southern Observatory, Karl-Schwarzschild-Strasse 2, 85748 Garching bei M\"{u}nchen, Germany }
\altaffiltext{28}{ SUPA School of Physics and Astronomy, University of St Andrews, North Haugh, St Andrews, KY16 9SS, Scotland, UK }
\altaffiltext{29}{ Max-Planck-Institut f{\"u}r Sonnensystemforschung, Max-Planck-Str. 2, 37191 Katlenburg-Lindau, Germany }
\altaffiltext{30}{ European Southern Observatory, Alonso de Cordova 3107,Vitacura, Santiago, Chile }
\altaffiltext{31}{ School of Mathematical Sciences, Queen Mary University of London, Mile End Road, London E1 4NS, England, UK }
\altaffiltext{32}{ University of Canterbury, Department of Physics \& Astronomy, Private Bag 4800, Christchurch, New Zealand}
\altaffiltext{33}{ Institut d'Astrophysique de Paris, Universit{\'e} Pierre \& Marie Curie, UMR7095 UPMC-CNRS 98 bis boulevard Arago, 75014 Paris, France }
\altaffiltext{34}{ European Southern Observatory, Casilla 19001, Vitacura 19, Santiago, Chile }
\altaffiltext{35}{ SOFIA Science Center, Mail stop N211-3, Moffett Field CA 94035, USA}
\altaffiltext{36}{ McDonald Observatory, 16120 St Hwy Spur 78, Fort Davis, TX 79734, USA}
\altaffiltext{37}{ Lawrence Livermore National Laboratory, IGPP, PO Box 808, Livermore, CA 94551, USA}
\altaffiltext{38}{ DSM/DAPNIA, CEA Saclay, 91191 Gif-sur-Yvette cedex, France}
\altaffiltext{39}{ Physics department, University of Rijeka, 51000 Rijeka, Croatia}
\altaffiltext{40}{ Technical University of Vienna, Dept. of Computing, Wiedner Hauptstrasse 10, Vienna, Austria}
\altaffiltext{41}{ Observatoire Midi-Pyrenees, UMR 5572, 14 avenue Edouard Belin, 31400 Toulouse, France}
\altaffiltext{42}{ Niels Bohr Institute, Astronomical Observatory, Juliane Maries Vej 30, 2100 Copenhagen, Denmark}
\altaffiltext{43}{ NASA Exoplanet Science Institute, Caltech, MS 100-22, 770 South Wilson Avenue Pasadena, CA 91125, USA}
\altaffiltext{44}{ Perth Observatory, Walnut Road, Bickley, Perth 6076, Australia}
\altaffiltext{45}{ South African Astronomical Observatory, PO Box 9 Observatory 7935, South Africa}
\altaffiltext{46}{ Space Telescope Science Institute, 3700 San Martin Drive, Baltimore, MD 21218, USA}
\altaffiltext{47}{ Astronomisches Rechen-Institut, Zentrum fur Astronomie, Heidelberg University, Monchhofstr. 12.14, 69120 Heidelberg, Germany}

\begin{abstract}
We report the extremely high magnification ($A > 1000$) binary
microlensing event OGLE-2007-BLG-514. We obtained good coverage around
the double peak structure in the light curve via follow-up
observations from different observatories.  
The binary lens model that includes the effects of parallax (known orbital motion of the Earth) 
and orbital motion of the lens yields a binary lens mass ratio of $q = 0.321 \pm 0.007$
and a projected separation of $s = 0.072 \pm 0.001$ in units of the Einstein radius.  
The parallax parameters allow us to determine the lens distance $D_L = 3.11 \pm 0.39$ kpc 
and total mass $M_L=1.40 \pm 0.18 M_\odot$; this leads to the primary and secondary components 
having masses of $M_1 = 1.06 \pm 0.13 M_\odot$ and $M_2 = 0.34 \pm 0.04 M_\odot$, respectively.
The parallax model indicates that the binary lens system is likely constructed by the main sequence stars. 
On the other hand, we used a Bayesian analysis to estimate probability distributions 
by the model that includes the effects of xallarap (possible orbital motion of the source around a companion) and parallax 
($q = 0.270 \pm 0.005$, $s = 0.083 \pm 0.001$). 
The primary component of the binary lens is relatively massive with 
$M_1 = 0.9_{-0.3}^{+4.6} M_\odot$ and it is at a distance of 
$D_{\rm L} = 2.6_{-0.9}^{+3.8}$ kpc. Given the secure mass ratio
measurement, the companion mass is therefore $M_2 = 0.2_{-0.1}^{+1.2} M_\odot$.  
The xallarap model implies that the primary lens is likely a stellar remnant, 
such as a white dwarf, a neutron star or a black hole. 

\end{abstract}


\section{Introduction}

Gravitational microlensing is one of the methods that can be employed
to search for planetary systems.  Current microlensing surveys focus
primarily on searching for extrasolar planets by monitoring dense star
fields in the direction of the Galactic bulge.  However, stellar
binary systems are also detectable using this method and their identification
is undoubtedly easier than for planetary binaries.
\citep{Udalski1994, Abe2003, Rattenbury2005, Hwang2011, Skowron2011, Shin2011}.

Stellar binary systems detected via high magnification microlensing
events have been studied by \citet{HanHwang2009} and
\citet{Shin2011}, and these events are particularly useful for
statistical studies of binary systems \citep{Gould2010, Shin2011}. 
This is because most high magnification candidates are detected before the
peak (or peaks) of the light curve and the high magnification regions
of the light curve are subsequently monitored by the follow-up
networks.  This usually leads to comprehensive coverage.  Therefore,
the physical details of the lens system will be revealed by the modeling.

Gravitational microlensing depends on the mass of the lens and is not
a function of lens brightness.  Therefore, microlensing can readily
detect faint stellar remnants such as white dwarfs, neutron stars and
black holes.  Due to their low luminosity, these objects are
difficult to detect. Isolated white dwarf stars are only really
observable optically in the nearby Galactic neighborhood, while
neutron stars are detected in the radio frequency domain as pulsars.
Stellar mass black holes have been discovered as part of binary
systems via X-ray and optical observations.  The MACHO collaboration
have interpreted several microlensing events in the direction of the
Galactic Bulge as due to isolated black holes with stellar masses:
MACHO-96-BLG-5, MACHO-98-BLG-6 \citep{Bennett2002} and MACHO-99-BLG-22 \citep{Agol2002}, 
which is the same event as OGLE-1999-BUL-32 \citep{Mao2002}. 
See also \citet{Poindexter2005}. 

In this paper, we report the high magnification microlensing event
OGLE-2007-BLG-514. We describe the
observations and data sets in Section \ref{sec:observation}.  The
light curve modeling with various effects is presented in Section
\ref{sec:modeling}.  We discuss the measurement of the source
magnitude and color, and derive the angular Einstein radius in Section
\ref{sec:correction}.  In Section \ref{sec:blending}, the blending
brightness for this event is discussed. 
The lens properties including mass and distance
are estimated in Section \ref{sec:properties}. Finally, we discuss our
results and conclusions in Section \ref{sec:conclusion}.

\section{Observations}
\label{sec:observation}

OGLE-2007-BLG-514 was announced as a likely microlensing event by the
Optical Gravitational Lensing Experiment (OGLE) Early Warning System
(EWS; \citealt{Udalski2003}) on 1 September 2007 (${\rm HJD'}
\equiv {\rm HJD} - 2,450,000 = 4345.62$) and independently by the
Microlensing Observations in Astrophysics (MOA) collaboration \citep{Bond2001} as
MOA-2007-BLG-464 on 26 September 2007.  The equatorial coordinates for
this event are $(\mathrm{R.A., Dec})_{\rm J2000.0} = (17^h 58^m03^s.09,~-27^\circ 31'05".7)$, 
which yields the Galactic coordinates $(l,b)=(2^\circ.62, -1^\circ.63)$.  
Forty days later (11 October), the OGLE collaboration issued a high-magnification alert.  
Two days after the high-magnification alert, the light curve reached its first peak allowing follow-up
observations to be conducted by ten telescopes all around the world.
  
% OGLE microlens alert mail: 2007.09.02 02:56 (JST), 2007.09.01 17:56 (UT)  
% OGLE High-mag  alert mail: 2007.10.11 19:01 (JST), 2007.10.11 10:01 (UT)
% MOA  microlens alert mail: 2007.09.26 09:45 (JST), 2007.09.26 00:45 (UT)
% JD=2454386.5 -> 2007.10.13

\begin{figure}\begin{center}
\begin{tabular}{cc}
\includegraphics[scale=0.6, angle=270, origin=b]{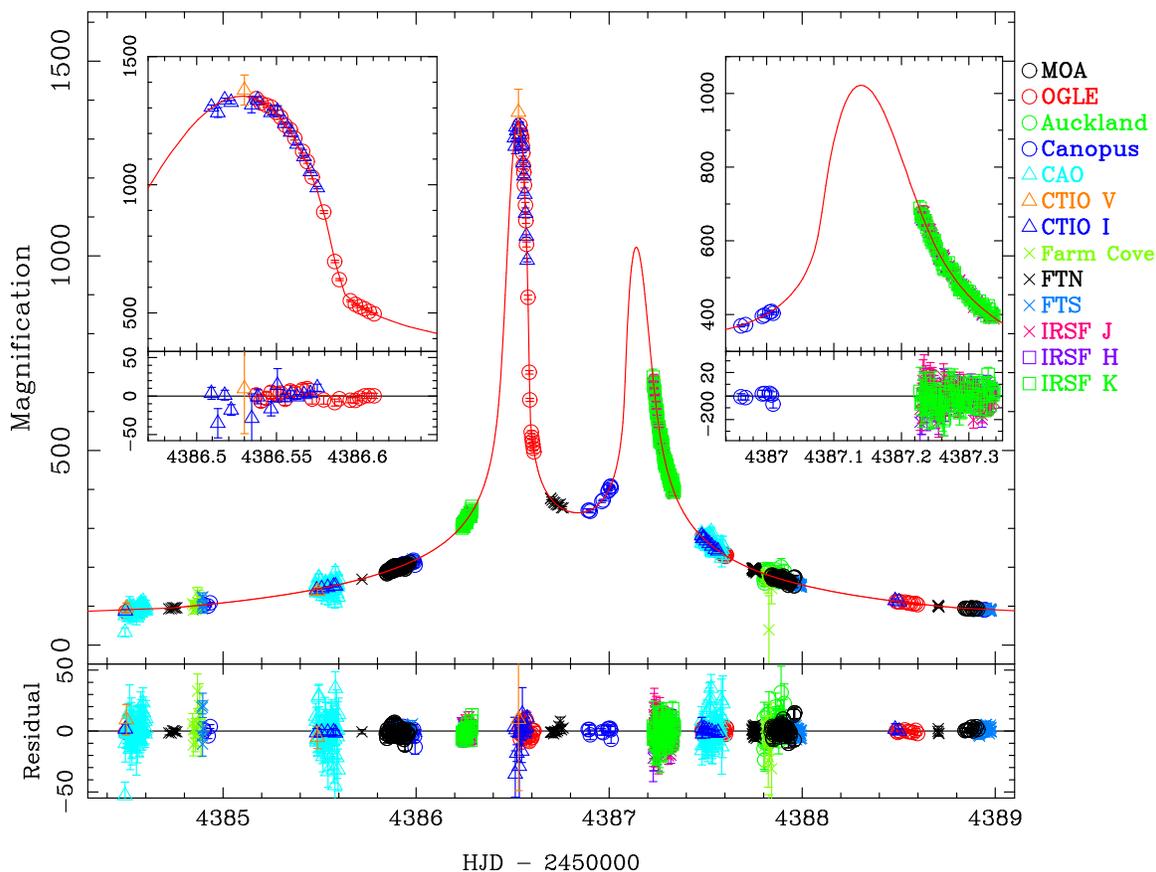} 
\end{tabular}
\caption{The light curve of the binary microlensing event, OGLE-2007-BLG-514. 
The top panel shows the data points and the best fit model light curve with both parallax and lens orbital motion effects. 
The residuals between the data points and the model are shown in the bottom panel. 
Close-up views of two peaks are shown in the corners. 
Note that the CTIO $H$-band data points have not been shown at the first peak of the light curve for good-looking, 
but we used in the modeling. }
\label{fig:lightcurve}
\end{center}\end{figure}

\begin{figure}\begin{center}
\includegraphics[scale=0.6, angle=270, origin=c]{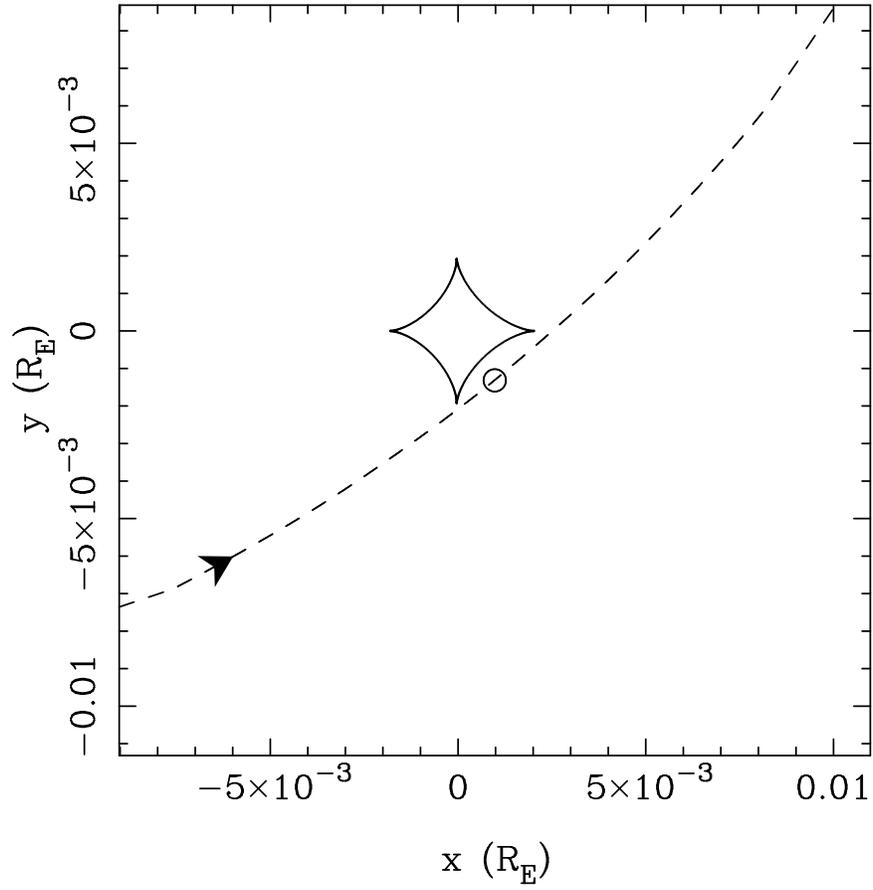} 
\caption{
The caustic curve (solid line) plotted for the OGLE-2007-BLG-514 best fit model. 
The dashed line indicates the source trajectory including both parallax and lens orbital motion effects. 
The circle on the source trajectory represents the source star size. }
\label{fig:caustic}
\end{center}\end{figure}

The double peak light curve for this event is shown in
Figure~\ref{fig:lightcurve}.  The first peak was covered by OGLE and
CTIO (HJD$'\sim$ 4386.5), the second peak was covered by Bronberg and
IRSF (HJD$'\sim$ 4387.2). The valley between the peaks was covered by
two telescopes, Canopus and Faulkes Telescope North. 
The caustic curve and source trajectory in the best-fit model is shown in Figure \ref{fig:caustic}.
The event was also observedusing high-resolution spectroscopy in order to identify the source
star properties \citep{Epstein2010,Bensby2010}.  We utilize this
information to estimate the limb darkening coefficients
(Section~\ref{sec:LD}).

The data set for OGLE-2007-BLG-514 consists of observations from 12
different observatories representing the OGLE, MOA, 
the Microlensing Follow Up Network ($\mu$FUN; \citealt{Yoo2004}), 
RoboNet \citep{Tsapras2009}, the Probing Lensing Anomalies
Network (PLANET; \citealt{Albrow1998}), 
as well as the InfraRed Survey Facility (IRSF).  
Specifically, the data set includes data
from the following telescopes, locations and passbands:
OGLE-III 1.3m Warsaw Telescope at Las Campanas Observatory (Chile) $I$-band, 
MOA-II 1.8m Telescope at Mt John University Observatory (New Zealand) wide-$R$-band, 
MOA 0.61m B$\&$C Telescope at Mt John University Observatory (New Zealand) $I$ and $V$ bands, 
$\mu$FUN 0.4m Telescope at Auckland Observatory (New Zealand) $R$-band, 
$\mu$FUN 0.35m Telescope at Bronberg Observatory (South Africa) unfiltered, 
$\mu$FUN 1.3m SMARTS Telescope at CTIO (Chile) $V$, $I$, and $H$ bands, 
$\mu$FUN 0.5m Telescope at Campo Catino Austral Observatory (CAO, Chile) unfiltered, 
$\mu$FUN 0.35m Telescope at Farm Cove Observatory (New Zealand) unfiltered, 
RoboNet 2.0m Faulkes Telescope North (FTN) at Haleakala Observatory (Hawaii) $R$ band, 
RoboNet 2.0m Faulkes Telescope South (FTS) at Siding Spring Observatory (Australia) $R$ band, 
PLANET 1.0m Telescope at Canopus Observatory (Australia) $I$ band, 
and IRSF 1.4m Telescope at SAAO (South Africa) $J$, $H$ and $K_{\rm S}$ bands. 

The various data sets were reduced using several methods. The OGLE data
were reduced using the standard OGLE DIA pipeline \citep{Udalski2003}.
The MOA, $\mu$FUN and IRSF data were reduced using the MOA DIA
pipeline \citep{Bond2001}.  $\mu$FUN CTIO $V,I$ band data were also
reduced using DoPHOT \citep{Schechter1993} which we used for the
source brightness and color in Section \ref{sec:correction}. 
The RoboNet data were reduced by the DanDIA pipeline \citep{Bramich2008}, 
and the PLANET data using pySIS version 3.0 reduction pipeline \citep{Albrow2009}.

The Bronberg unfiltered data were not included in the modeling because
it may require airmass corrections. Fortunately, IRSF $J,H,K_{\rm s}$-band 
data covered the same region as the Bronberg data. We have
checked that the result was not affected by whether the Bronberg data
were included or not.  The MOA B\&C data were also not included in
the modeling because all of the three day's worth of data points are
poor quality due to the weather conditions.  Moreover, the same region was
covered by the MOA-II data.  Both telescopes are located at Mt John, 
and the MOA-II telescope yields better quality data.

The error bars for the data points have been re-normalized such that the 
reduced $\chi^2$ of the best fit model $\chi^2/d.o.f. \simeq 1$.
We used the formula,
\begin{equation}
\sigma'_{i} = k \sqrt{\sigma_i^2 + e_{\rm min}^2}, 
\end{equation}
where $\sigma_i$ is the original error value of the $i$th data point
in magnitude units, and the re-normalizing parameters are $k$ and
$e_{\rm min}$.  This non-linear formula operates so that the error bars at
high magnification are affected by the parameter $e_{\rm min}$, 
which can be described flat-fielding errors. 
For the parameter of $e_{\rm min}$, 
we plot the cumulative distribution of $\chi^2$ where the data points were sorted by error in magnitude, 
and we choose values of $e_{\rm min}$ such that the cumulative distribution is a straight line with slope of 1. 
Then, the parameter of $k$ is chosen to be $\chi^2/d.o.f. \simeq 1$. 
The $k$ and $e_{\rm min}$ are determined separately for each data set and are shown in Table
\ref{tb:norm}.

\section{Modeling}
\label{sec:modeling}

It is clear from Figure \ref{fig:lightcurve} that OGLE-2007-BLG-514 is
not a single-lens event due to the double peak structure in the light
curve.  From the light curve shape alone we can deduce that either the
source passed in close proximity to two caustic cusps or it passed
across the caustic itself.  Moreover, the extremely high magnification
($A > 1000$) of the event indicates that the source star passed
near the central caustic.  Since the OGLE and CTIO telescopes covered
the first peak of the light curve, finite source effects must be included 
in the modeling.  Thus, our strategy began by searching for best fit
parameters using a binary lens model and finite source parameters. We
subsequently introduced parameters to incorporate higher order
effects.

\subsection{Limb Darkening}
\label{sec:LD}

To properly model finite source effects we need to allow for limb
darkening, which accounts for the changing brightness between the
source disk center and the rim.  
We adopted a linear limb darkening law 
with one parameter for the source brightness
\begin{equation}
S_\lambda (\vartheta ) = S_\lambda (0)\left[ 1-u(1-\cos\vartheta )\right]. 
\label{eq:LD}
\end{equation}
Here, $u$ is the limb darkening coefficients, $S_\lambda (0)$ is 
the central surface brightness of the source, and $\vartheta$ is the angle 
between the normal to the stellar surface and the line of sight, i.e., 
$\sin \vartheta$=$\theta$/$\theta_*$, 
where $\theta$ is the angular distance from the center of the source as measured in the plane of the sky.

From \citet{Bensby2010}, the source star (OGLE-2007-BLG-514S) is a G dwarf in the Galactic bulge, 
$T_{\rm eff}$ = 5644 $\pm$ 130 K, $\log g$ = 4.10 $\pm$ 0.28 cm s$^{-2}$, 
and high-metallicity log[Fe/H] = 0.27 $\pm$ 0.09 dex. 
Therefore, we fix the limb darkening coefficients selected from \citet{Claret2000} 
with effective temperature $T_{\rm eff}$ = 5500 K, surface gravity $\log g$ = 4.0 cm s$^{-2}$ 
and metallicity $\log$[M/H] = 0.3 (Table~\ref{tb:LDparameter}).

\subsection{Binary lens model}
\label{sec:binary}

The best fit binary lens model parameters were searched for using a
Markov Chain Monte Carlo (MCMC) approach that frequently changes 
the ``jump function" in order to efficiently locate the minimum $\chi^2$ value.
For example, refer to \citet{Verde2003}, \citet{DoranMueller2004} and
\citet{Bennett2010}. A single lens microlensing model has three
parameters: the time of peak magnification, $t_0$, the Einstein radius
crossing time, $t_{\rm E}$, and the minimum impact parameter, $u_0$.
A binary lens model requires three additional parameters: the mass
ratio, $q$, which is the mass of the companion relative to the mass of
the primary lens; the binary lens separation, $s$, which is the
separation of the binary components projected on to the lens plane and
normalized to the Einstein radius; and the angle of the source
trajectory relative to the binary lens axis, $\alpha$.  An additional
required parameter, $\rho$, is the source radius relative to the
angular Einstein radius which 
in combination with
the limb darkening law is
used to model the finite source effects.  Furthermore, there are two
parameters for each data set and passband that are required to
describe the individual unmagnified source and background fluxes.

We conducted a broad parameter search with initial parameters
distributed over the ranges $-3<\log q<0$ and $-2<\log s<1$.  The best
models from this search involved a binary lens with a mass ratio of $q
\geq 0.1$ and the trajectory of the source star making a close
approach to the central caustic (Figure \ref{fig:caustic}).  Given
that the primary lens has a stellar mass, the lens companion is also a
star with a stellar mass and not a planet.  Two models with
approximately equal $\chi^2$ values yielded lens component separations
of $s=0.08$ and $s=17$ in units of the Einstein radius.  This degeneracy in
the component separation $s\leftrightarrow s^{-1}$ for a central
caustic event was predicted by \citet{Dominik1999}.

The central caustic has a diamond-like shape. 
So, we expect that the source
trajectory angle $\alpha$ will have four possible solutions involving
approaches to two caustic cusps.  The general parameter
search yielded two preferred solutions with $\alpha = 0.7$ and
$5.6$, 
which correspond to the binary degeneracy \citep{Skowron2011}. 
The other two expected angles $\alpha = 2.5$ and $3.8$ do
not generate good models compared with the $\alpha = 0.7$ and
$5.6$ models.  
This is because the magnification contours near the caustic are not perfectly symmetric.
As a result, we get four binary lens
models with close/wide separation degeneracies and impact parameter
$u_0$ degeneracies.  
The best fit parameters are listed in Table \ref{tb:parallax}, and are denoted as close1 and wide1.

\subsection{Parallax Effect}
\label{sec:parallax}

From the binary modeling (Section \ref{sec:binary}), the Einstein radius crossing time, $t_{\rm E}$, is very long,  
$t_{\rm E} > 200$ days, implying that 
there is a good chance to detect Earth's orbital parallax effects in the light curve. 
Microlens parallax falls into two different categories, the Earth's orbital motion around the Sun 
and the difference of telescope location on the Earth. 
These parallaxes are called ``orbital parallax" and ``terrestrial parallax", respectively. 
We searched for a parallax model including these effects both separately (models ``2" and ``3" in the Table \ref{tb:parallax}) 
and combined (models ``4" in the Table \ref{tb:parallax}). 
From now on models with ``parallax" will mean with both effects taken into account. 
Two additional parameters, $\pi_{\rm E,N}$ and $\pi_{\rm E,E}$, express the parallax effect on the light curve \citep{Gould2000}. 
These are the two components of the microlens parallax amplitude, $\pi_{\rm E} = \sqrt{\pi_{\rm E,N}^2 + \pi_{\rm E,E}^2}$. 
The parallax amplitude $\pi_{\rm E}$ is also represented by the lens-source relative parallax, 
$\pi_{\rm rel} = \pi_{\rm L} - \pi_{\rm S}$, and the angular Einstein radius $\theta_{\rm E}$,
\begin{equation}
\pi_{\rm E} = \frac{\pi_{\rm rel}}{\theta_{\rm E}} = \frac{\rm AU}{\tilde{r}_{\rm E}}, 
\label{eq:parallax}
\end{equation}
where $\tilde{r}_{\rm E}$ is the Einstein radius projected onto the observer plane. 
From the parallax amplitude parameter, the degeneracy of three physical parameters: 
lens mass, distance and transverse velocity in $t_{\rm E}$, is broken, allowing us to determine the properties of the lens. 

In our parallax model search, we explored four classes of models with close and wide separations  
and also in case of impact parameter $u_0>0$ and $u_0<0$ based on the results of the binary lens model search. 
The parameters for the parallax model are listed in Table \ref{tb:parallax}. 
We found that the $\chi^2$ of the best fit parallax model was improved by 19 over the non-parallax model, 
and these models are denoted as ``4" in the Table. 
Most of the improvement of $\chi^2$ is from the MOA and OGLE data. 
This is reasonable because the MOA and OGLE data cover a large portion of the light curve 
rendering orbital parallax effects significant for the light curve. 
The $u_0>0$ model is a better fit than the $u_0<0$ model, but 
since the improvement is only $\Delta \chi^2 = -8$, the $u_0$ sign degeneracy is not decisively resolved. 

\subsection{Orbital motion of the lens}

Aside from the microlens parallax effects, it is possible that the lens orbital motion effect also have influenced the light curve.
The orbital motion of the lens has strong effects when the source passes through or approaches the caustics. 
For the orbital motion of the lens, we require two additional orbital motion parameters, $\omega$ and $ds/dt$, 
\citep{An2002}. These parameters indicate the binary rotation rate, and the uniform expansion rate in binary separation, $s$. Therefore, the new ones of $\alpha'$ and $s'$ are described as
\begin{equation}
\alpha' = \alpha + \omega (t-t_0), \hspace{20pt} s' = s + ds/dt (t-t_0). 
\end{equation}

The results of the lens orbital motion modeling are shown in Table \ref{tb:OM}. 
The close separation model is better fit than the wide separation model. 
But, the $u_0$ sign degeneracy is not decisively resolved. 
Then, we also performed fits including both parallax and lens orbital motion effects. 
The results are shown in Table \ref{tb:OM}.
The best fit model with the parallax and lens orbital motion indicates that 
the companion of the lens have the orbital motion parameters of 
$\omega = (-4.857 \pm 0.004) \times 10^{-2}$ rad days$^{-1}$ and $ds/dt = (9.4 \pm 0.1) \times 10^{-4}$ days$^{-1}$. 
It is already known that $\omega$ is often degenerate with $\pi_{E, \perp}$,
which is the component of $\pi_E$ perpendicular to
the apparent acceleration of the Sun projected on the sky \citep{Batista2011, Skowron2011}.
For this event, we could break this degeneracy. 

For composition of the lens system, the projected velocity of the lens companion should be smaller than 
the escape velocity of the lens system: $v_\perp \leq v_{\rm esc}$ \citep{An2002}, where, 
\begin{eqnarray}
v_\perp &=& \sqrt{(ds/dt)^2 + (\omega s)^2} D_{\rm l} \theta_{\rm E}, \\
v_{\rm esc} &=& \sqrt{\frac{2GM}{r}} \leq v_{\rm esc,\perp} = \sqrt{\frac{2GM}{r_\perp}}, 
\end{eqnarray}
and where $r_\perp = s\theta_{\rm E} D_{\rm L}$.
We confirmed that the results for each model was not over the escape velocity of the lens system, 
using the lens mass and distance calculated by the combined parallax and lens orbital motion model (Section \ref{sec:properties}).

\subsection{Xallarap Effect}
\label{xallarap}

The orbital parallax effect is caused by the Earth's orbit around the Sun. On the other hand, 
if a companion star is orbiting about the source star, the light curve is affected in the same way with microlens parallax. 
This effect is called ``xallarap". It has been discussed that 
the Earth's orbital parallax effect can be degenerate with the xallarap effect \citep{Poindexter2005}. 
The xallarap model has five additional parameters: the two components of xallarap amplitude, 
$\xi_{\rm E,N}$ and $\xi_{\rm E,E}$, which represent the xallarap amplitude, $\xi_{\rm E} = \sqrt{\xi_{\rm E,N}^2 + \xi_{\rm E,E}^2}$,  
the direction of observer relative to the source orbital axis, R.A.$_\xi$ and decl.$_\xi$ , and the orbital period, $P_\xi$.
For an elliptical orbit, two additional parameters are required: 
the orbital eccentricity, $\epsilon$ and time of periastron, $t_{\rm peri}$.
In our xallarap model fit, the two parameters for an elliptical orbit are fixed as the parameters for the Earth's orbit. 
Note that we assumed that the brightness from the companion of the source is low, so we did not include the additional parameter for the source companion brightness. 

We also impose on our xallarap the Kepler constraint. The $\xi_{\rm E}$ is represented by the following equation, estimated from Kepler's third law: 
\begin{equation}
\xi_{\rm E} = \frac{a_{\rm s}}{\hat{r}_{\rm E}} = \frac{1{\rm AU}}{\hat{r}_{\rm E}}\frac{M_{\rm c}}{M_{\odot}}\Biggl(\frac{M_{\odot}}{M_{\rm s}+M_{\rm c}} \frac{P_{\xi}}{1{\rm yr}}\Biggr)^{\frac{2}{3}},
\label{eq:xallarap_amp}
\end{equation}
where $\hat{r}_{\rm E}$ is the Einstein radius projected onto the source plane 
($\hat{r}_{\rm E}=\theta_{\rm E} D_{\rm S}$), $a_{\rm s}$ is the separation of the source companion, 
$M_{\rm s}$ and $M_{\rm c}$ are the mass of the primary and companion source, respectively.
To find the best xallarap model that is allowed by Kepler's third law, we have done MCMC runs with an
additional constraint to $\chi^2$ \citep{Sumi2010} given by 
\begin{equation}
\chi_{\rm orb}^2 = \Theta (\xi_{\rm E,max}-\xi_{\rm E}) \Biggl( \frac{\xi_{\rm E,max}-\xi_{\rm E} }{\sigma_{\xi_{\rm E,max}}} \Biggr)^2,
\label{eq:xallarap_chi2}
\end{equation}
where $\xi_{\rm E,max}$ is evaluated by Equation \ref{eq:xallarap_amp} with parameters in
each step of the MCMC and fixed values of $M_{\rm s}=M_{\rm c}=1M_{\odot}$
and 50\% error in $\xi_{\rm E,max}$, which depend only weakly on other parameters. 
Here, $\Theta$ is the Heaviside step function.

\begin{figure}\begin{center}
\includegraphics[scale=0.7, angle=270, origin=b]{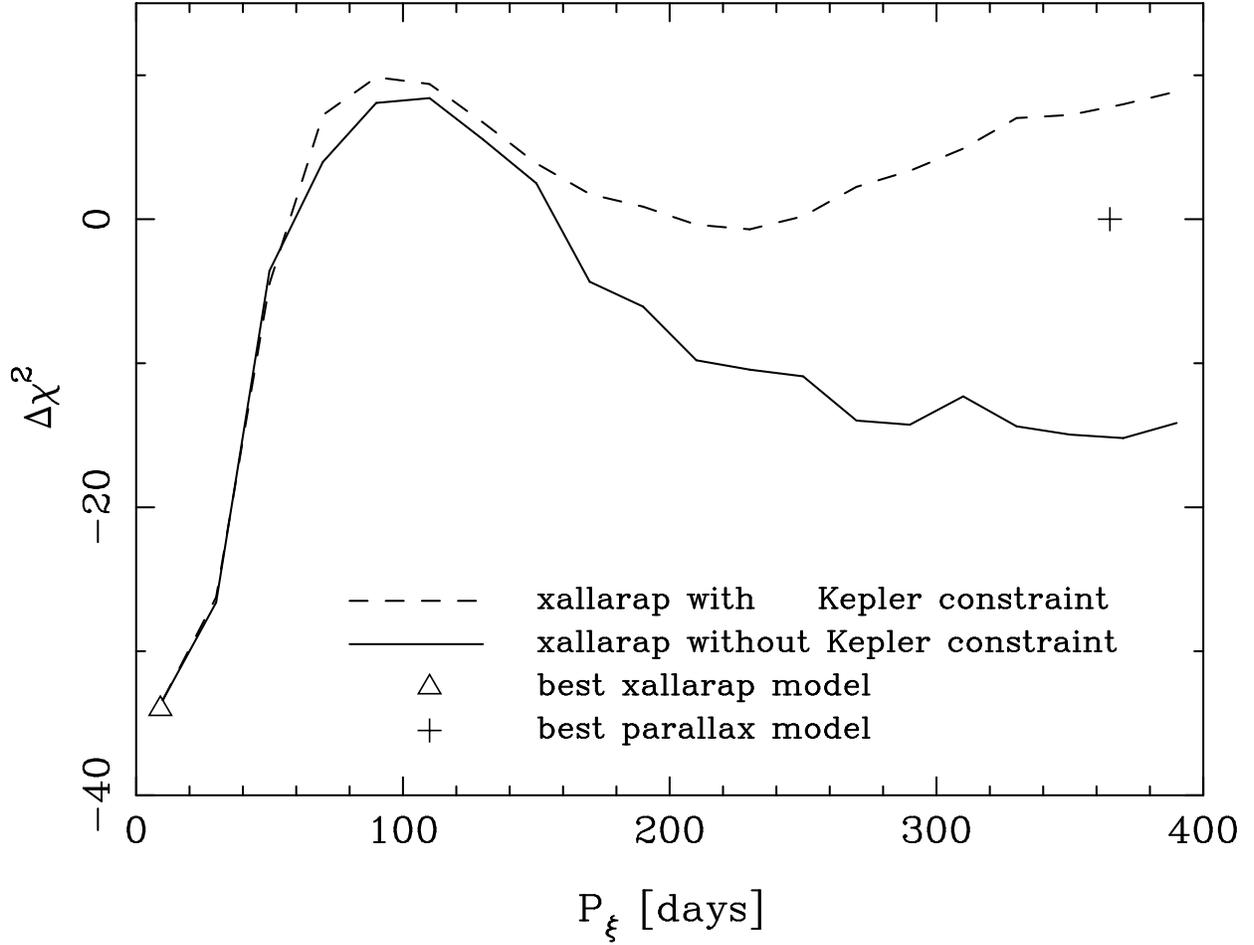} 
\caption{The $\Delta \chi^2$ of the best xallarap model as a function of the orbital period of the
source star and its companion. The dashed and solid lines indicate 
the model with and without the Kepler constraint, respectively. The cross represents the best fit model with parallax effect. }
\label{fig:xallarap}
\end{center}\end{figure}

First, we prepared the initial parameters of the orbital period as fixed parameters in the fits 
in interval of 20 days from 10 to 400 days, and the others as free parameters. 
The $\chi^2$ distribution for the orbital period is shown in Figure \ref{fig:xallarap}. 
We found that a short orbital period produces a good model, so we restarted the fit around $P_\xi =10$ days 
without any fixed parameters. 
Note that the result is not changed even if the model includes the Kepler constraint, 
due to the small xallarap amplitude (which is shown by the dashed line in Figure \ref{fig:xallarap}). 

The best fit xallarap model has $\chi^2= 3545$, which is significantly better than the only parallax effect model, 
giving $\Delta \chi^2 = (\chi^2_{\rm xallarap} - \chi^2_{\rm parallax}) = -34$. 
The best fit parameters are shown in Table \ref{tb:xallarap}. 
The parameter $P_\xi$ is not consistent with the Earth's orbital period, so this xallarap model is different than the parallax model.

\subsection{Modeling Summary} 
\label{sec:modelingsummary} 

The best fit xallarap $\chi^2$ value is smaller than the parallax model and the combined parallax and lens orbital motion model. 
Moreover, the xallarap model is different than the parallax model. 
This implies that the xallarap effect is considered more significant than the parallax effect on the light curve. 
However, spurious xallarap signals at this level are common in microlensing \citep{Poindexter2005}. 
And, the data on this event has high Signal-to-noise ratio (S/N), so we expect that stronger false signals are caused by systematics. 
Therefore, the parallax and lens orbital motion model is the most plausible solution, 
but we can not absolutely excluded the xallarap model. 
Thus, we consider two solutions of the parallax model and the xallarap model for the estimation of the lens properties. 
For the estimation of the angular Einstein radius, $\theta_{\rm E}$, in Section \ref{sec:correction}, 
we used the best fit parameters of the model with both parallax and lens orbital motion effects 
(close separation and the $u_0<0$ model). 
The best fit model has a binary mass ratio of $q = 0.321 \pm 0.007$ and 
a separation of $s = 0.072 \pm 0.001$ in a unit of Einstein radius 
with $\pi_{\rm E} = 0.13 \pm 0.02$.  

\begin{figure}\begin{center}
\includegraphics[scale=0.7, angle=0, origin=c]{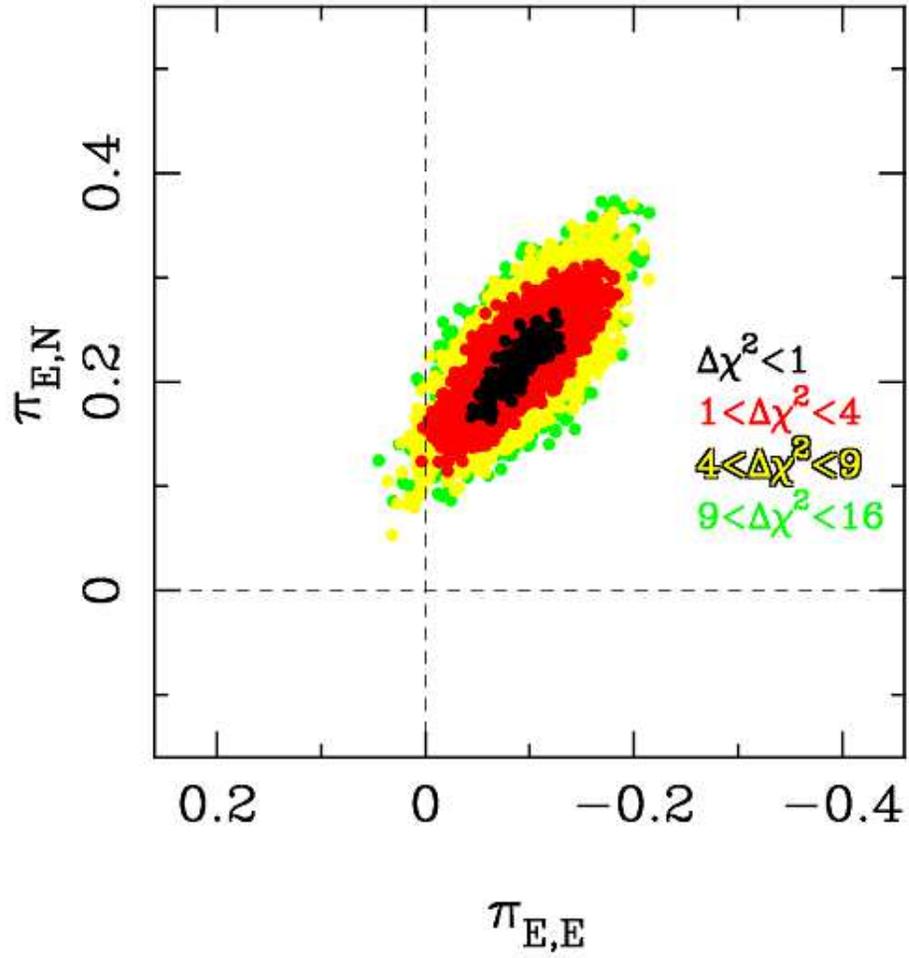} 
\caption{
$\Delta \chi^2$ map of the parallax parameters of the best fit model with xallarap and parallax effects. 
The colors of black, red, yellow, green represent $\Delta \chi^2 = 1, 4, 9, 16$, respectively. }
\label{fig:parallax}
\end{center}\end{figure}

In contrast, the probability distribution of the lens properties was also estimated 
by using the xallarap model parameters in Section \ref{sec:properties}. 
We performed fits including both parallax and xallarap effects 
because parallax amplitude is related to the lens mass. 
The results are listed in Table \ref{tb:xallarap}. 
The xallarap and parallax model has a binary mass ratio of $q = 0.270 \pm 0.005$ and 
a separation of $s = 0.083 \pm 0.001$ in a unit of Einstein radius 
with $\xi_{\rm E} = (3.91 \pm 0.18) \times 10^{-4}$, and $P_\xi = 9.139 \pm 0.006$ days. 
Since the orbital period of the source companion is very short, 
it may be that parallax and xallarap do not affect the same part of the light curve. 
It is possible that the parallax results are unaffected by xallarap effects. 
Therefore, the parallax signal has not been detected on the light curve significantly,  
so we set the upper limit of the parallax amplitude $\pi_{\rm E} < 0.5$ 
from the $\Delta \chi^2$ contour map of parallax parameters in Figure \ref{fig:parallax}.  
This upper limit will imply a lower limit of the lens mass. 

\section{Source star and the Angular Einstein Radius}
\label{sec:correction}

\begin{figure}\begin{center}
\includegraphics[scale=0.7, angle=270, origin=c]{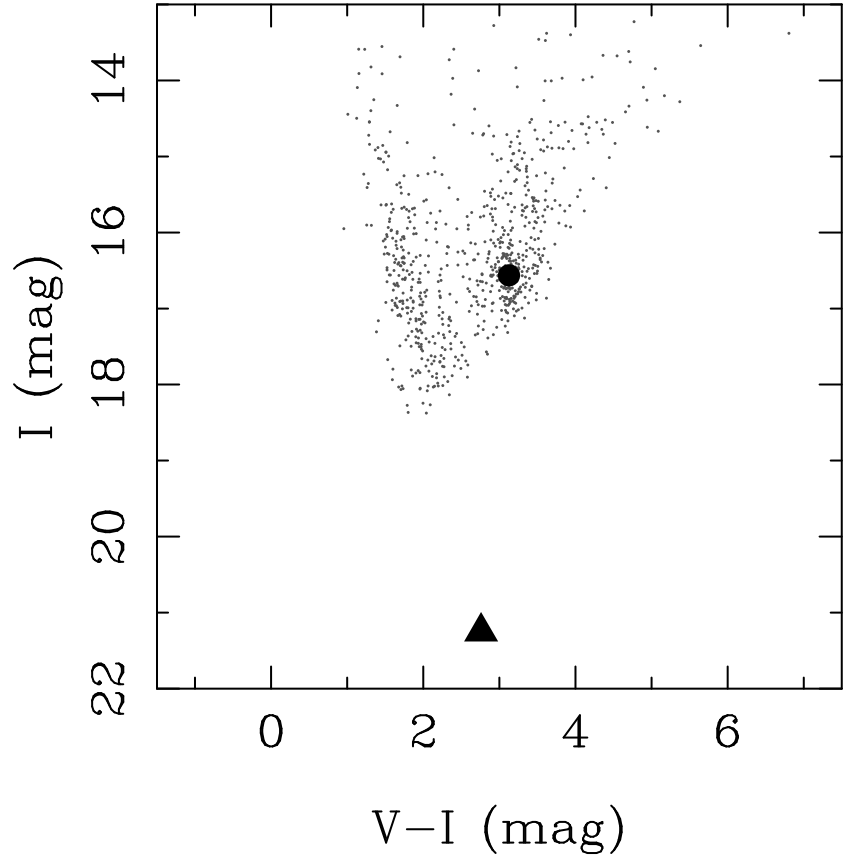} 
\caption{The $(V-I, I)$ color magnitude diagram of stars within 2$'$ 
of the OGLE-2007-BLG-514 source using $\mu$FUN CTIO data calibrated to the OGLE-III catalog. 
The triangle and dot indicate the source star and the center of the
red clump giant, respectively.}
\label{fig:cmd}
\end{center}\end{figure}

The source star angular radius $\theta_*$ was determined using source magnitude and color. 
The source star magnitudes and colors estimated from the light curve fit need to be corrected for extinction and reddening due to the interstellar dust in the line of sight.
The Red Clump Giant (RCG) is the standard candle to estimate extinction and reddening. 
The Color Magnitude Diagram (CMD) was made from CTIO $I$- and $V$- band stars within 2$'$ of the source star 
calibrated to the OGLE-III calalog (Figure~\ref{fig:cmd}). 
In this CMD, we find the RCG centroid to be, 
\begin{equation}
(I, V-I)_{\rm clump, obs} = (16.56, 3.13) \pm (0.03, 0.03).
\end{equation}

We adopt the Galactic bulge RCG magnitude $M_{I,RC,0} = -0.10 \pm 0.05$ from \citet{Nataf2012} 
and color $(V-I)_{RC,0} = 1.06 \pm 0.06$ from \citet{Bensby2011}. 
According to \citet{Nishiyama2005}, the clump in this field is 0.1 mag brighter than the Galactic center, 
which we take to be at $R_{\rm 0} = 8.0 \pm 0.3$ kpc \citet{Yelda2010}. 
Hence, the distance modulus of the clump is DM=$14.42 \pm 0.09$. 
Thus, the dereddened RCG centroid is expected to be
\begin{equation}
(I, V-I)_{\rm clump, 0} = (14.32, 1.06) \pm (0.11, 0.06). 
\end{equation}
Comparing these centroids, we find the extinction value $A_{I}$ and reddening value $E(V-I)$ to be,
($A_{I}, E(V-I)$) = (2.24, 2.07) $\pm$ (0.11, 0.06). 

The source star magnitudes from the light curve fit are $(I, V-I)_{\rm s, obs} = (21.44, 2.76) \pm (0.01, 0.03)$. 
Applying $A_{\rm I}$ and $E(V-I)$, the dereddended source color and magnitude $(I, V-I)_{s,0}$ is calculated to be 
\begin{equation}
(I, V-I)_{\rm s, 0} = (19.20, 0.69) \pm (0.11, 0.07). 
\end{equation}

We estimate $(V-K)_{\rm S, 0} = 1.50 \pm 0.18$ magnitude from $(V-I)_{\rm S, 0}$ and \citet{Bessell1988} color-color relation, 
and so $K = 18.40 \pm 0.22$.  
From \citet{Kervella2004}, the relationship between color/brightness and a star angular radius is   
\begin{equation}
\log(2\theta_*) = 0.0755(V-K) + 0.5170 - 0.2K, 
\label{eq:source radius}
\end{equation}
so the source angular radius is
\begin{equation}
\theta_* = 0.45\pm 0.02 \hspace{0.1cm} \mu \mathrm{as}. 
\end{equation}
Thus, the angular Einstein radius $\theta_{\rm E}$ is calculated by 
the source angular radius $\theta_*$ and $\rho = (2.98 \pm 0.04) \times 10^{-4}$ to be 
\begin{equation}
\theta_E = \frac{\theta_*}{\rho} = 1.50 \pm 0.08 \hspace{0.1cm} \mathrm{mas}, 
\end{equation}
and finally the lens-source relative proper motion $\mu_{\rm rel}$ is 
\begin{equation}
\mu_{\rm rel} = \frac{\theta_E}{t_{\rm E}} = 2.70 \pm 0.15 \hspace{0.1cm} \mathrm{mas} \hspace{0.1cm} \mathrm{yr^{-1}}.
\end{equation}

\section{Blended Light} 
\label{sec:blending} 

Blending magnitudes obtained from the fits are listed in Table~\ref{tb:flux}. 
The result of the fit indicates that the blending flux in the CTIO $V$-band is $-16.9 \pm 8.7$ ADU, 
however the calculated flux is consistent with zero flux to within 2 sigma. 
The blending flux in the CTIO $I$-band is also consistent with zero flux to within 2 sigma. 
Thus, we estimate the upper limit of the brightness by using baseline images 
from the CTIO $I,V$-band and IRSF $J,H,K_{\rm s}$-band, by taken about 4 years after the peak 
so that the source star is not magnified by the microlensing. 
The $I$- and $V$-band magnitude were calibrated to the OGLE-III catalog magnitude, 
and the $J,H,K_{\rm s}$ band magnitude were calibrated to the 2MASS catalog. 
The result is that $V>20.90$, and other results are listed in Table~\ref{tb:flux}. 
These upper limits are estimated from the equation, 
$M_{\rm limit} = -2.5 \log{(5\sqrt{f_{\rm sky} \times \pi r^2/G})} + c$,
where $f_{\rm sky}$ is sky flux at this target position, $\pi r^2$ is a PSF area, $G$ is the gain value,
$c$ is the scale factor to calibrate to catalog magnitude. 
These parameters are listed in Table \ref{tb:upperlimit}. 
For $I$-band upper limit on the brightness, we use the results of the modeling in the CTIO data to estimate, $I>19.15$.

\section{The Lens Properties}
\label{sec:properties}

In Section \ref{sec:modeling} and \ref{sec:correction}, we obtained the best fit model parameters and 
the angular Einstein radius from the source brightness. 
From them, we estimated the lens properties for the two solutions of the parallax model and the xallarap model.  

\subsection{The parallax model}

The both of finite source and parallax effects allow us to determine the physical parameters of the lens properties 
as the lens total mass $M_{\rm L}$, distance $D_{\rm L}$ and velocity $v$. 
The lens total mass is described by the angular Einstein radius $\theta_{\rm E}$ and 
microlens parallax amplitude $\pi_{\rm E}$ as given by 
\begin{eqnarray}
M &=& \frac{\theta_{\rm E}}{\kappa \pi_{\rm E}}, 
\label{eq:mass}
\end{eqnarray}
where $\kappa = 4G/(c^2~{\rm AU}) = 8.144~{\rm mas}~M_{\odot}^{-1}$. The distance is derived from the Equation \ref{eq:parallax}. 

We calculate the lens properties from the parameters of the combined parallax and lens orbital motion model
(close separation and $u_0<0$). 
The lens total mass is $M_{\rm L}=1.40 \pm 0.18 M_\odot$, as the components of the lens, the primary lens has 
$M_1 = 1.06 \pm 0.13  M_\odot$ with a companion $M_2 = 0.34 \pm 0.04 M_\odot$. 
The distance to the lens is $D_{\rm L} = 3.11 \pm 0.39~\mathrm{kpc}$. 

The upper limit of the brightness derives the upper limit of the lens mass. 
The absolute magnitude $M_I$ is calculated by 
\begin{eqnarray}
M_I &=& I_{\rm b} - 5 \log{\frac{D_{\rm L}}{10 {\rm pc}}} -A_{I,{\rm L}}, \\
    &=& I_{\rm b} - 5 \log{\frac{D_{\rm L}}{10 {\rm pc}}} -A_{I,{\rm S}} + (A_{I,{\rm S}}-A_{I,{\rm L}}), 
\end{eqnarray}
where $I_{\rm b}$ is the apparent magnitude of the blend, $A_{I,\rm S}$ and $A_{I,\rm L}$ are the extinction 
toward the source and lens, respectively. 
Since the lens should be in front of the source, the extinction toward the source is larger than the one of the lens, 
($A_{I,\rm S}-A_{I,\rm L})>0$. 
Therefore, the upper limit of the brightness is $M_I > 4.4$. 
Note that this upper limit is quite conservative because there is a possibility that some of the dust is behind the lens. 
The upper limit of the brightness implies that the upper limit of the lens mass is $M < 0.99 M_\odot$, 
assuming the primary lens is a main sequence star. 
The main sequence star is the dominant component in the galaxy, 
and the upper limit of the lens mass, assuming that the primary lens is a main sequence star, 
is consistent within the error bars of the primary lens mass 
calculated by the parallax model parameters.
Thus, the parallax model indicates that the binary lens system is likely constructed by the main sequence stars.

\subsection{The xallarap model}

We also estimated the probability distribution of the lens properties using the combined xallarap and parallax model 
(close separation and $u_0>0$). 
From the fit parameter for the finite source effects, $\rho$,
we were able to calculate the angular Einstein radius.  
In Section \ref{sec:correction}, the source angular radius, $\theta_*$, and the angular Einstein radius, $\theta_{\rm E}$, are estimated 
by the combined parallax and lens orbital motion model. 
On the other hand, for the combined xallarap and parallax model,  
the $f_{\rm S} \times t_{\rm E}$ were invariant with the different model. Therefore,  
we can estimate these quantities using the equation \citep{Yee2009}, 
\begin{equation}
\theta_* = \frac{\sqrt{f_{\rm S}}}{Z},
\end{equation}
where $f_{\rm S}$ is the source flux as determined from the model, and $Z$ is the remaining set of factors. 
The $\theta_{\rm *,xa}$ and $\theta_{\rm E,xa}$ are calculated by the model parameters of 
$t_{\rm E,xa} = 169.4 \pm 1.0$ days and $\rho_{\rm xa} = (3.45 \pm 0.02) \times 10^{-4}$ that 
$\theta_{\rm *, xa} = \theta_* \times \sqrt{f_{\rm S,xa}/f_{\rm S}} = \theta_* \times \sqrt{t_{\rm E}/t_{\rm E,xa}} = 0.49 \pm 0.03$, 
and  $\theta_{\rm E, xa} = \theta_{\rm *,xa}/\rho_{\rm xa} = 1.42 \pm 0.10$ mas. 
Note that the index of ``xa" indicates the parameter of the combined xallarap and parallax model.
We estimate the probability distribution from Bayesian analysis 
by combining the Equation \ref{eq:mass} and the measured values of $\theta_{\rm E,xa}$ and $t_{\rm E,xa}$.  

The mass function adopted for the calculation based on \citet{Sumi2011} Table~S3, model~\#1. 
The main sequence stars and brown dwarf stars form a power-law function, $dN/dM = M^{-\alpha}$ where 
$\alpha = 2.0$ for ($ 0.7  < M/M_\odot < 1.0 $), 
$\alpha = 1.3$ for ($0.08  < M/M_\odot < 0.7 $) and $\alpha=0.5$ for ($0.01 < M/M_\odot < 0.08 $). 
The stellar remnant stars are assumed to be white dwarfs (WDs; $M = 0.6 M_\odot, \sigma = 0.16$), 
neutron stars (NSs; $M = 1.35 M_\odot, \sigma = 0.04$) and black holes (BHs; $M = 5 M_\odot, \sigma = 1$). 
These remnants are represented by Gaussians. 
The fraction of initial numbers of these objects in the four classes MSs (which include main sequence and brown dwarf stars), 
WDs, NSs and BHs are distributed as 88.8, 10.0, 1.0, and 0.2, respectively. 
The distance to the Galactic center is assumed to be 8 kpc and 
the upper limit of microlens parallax amplitude, $\pi_{\rm E}<0.5$, is adopted in the calculation, 
which affects the lower limit of the lens mass. 
The upper limit of the brightness, $I_{\rm b,0} = I_{\rm b} - A_I > 16.91$, is also included. 

\begin{figure}\begin{center}
\includegraphics[scale=0.65, angle=270, origin=b]{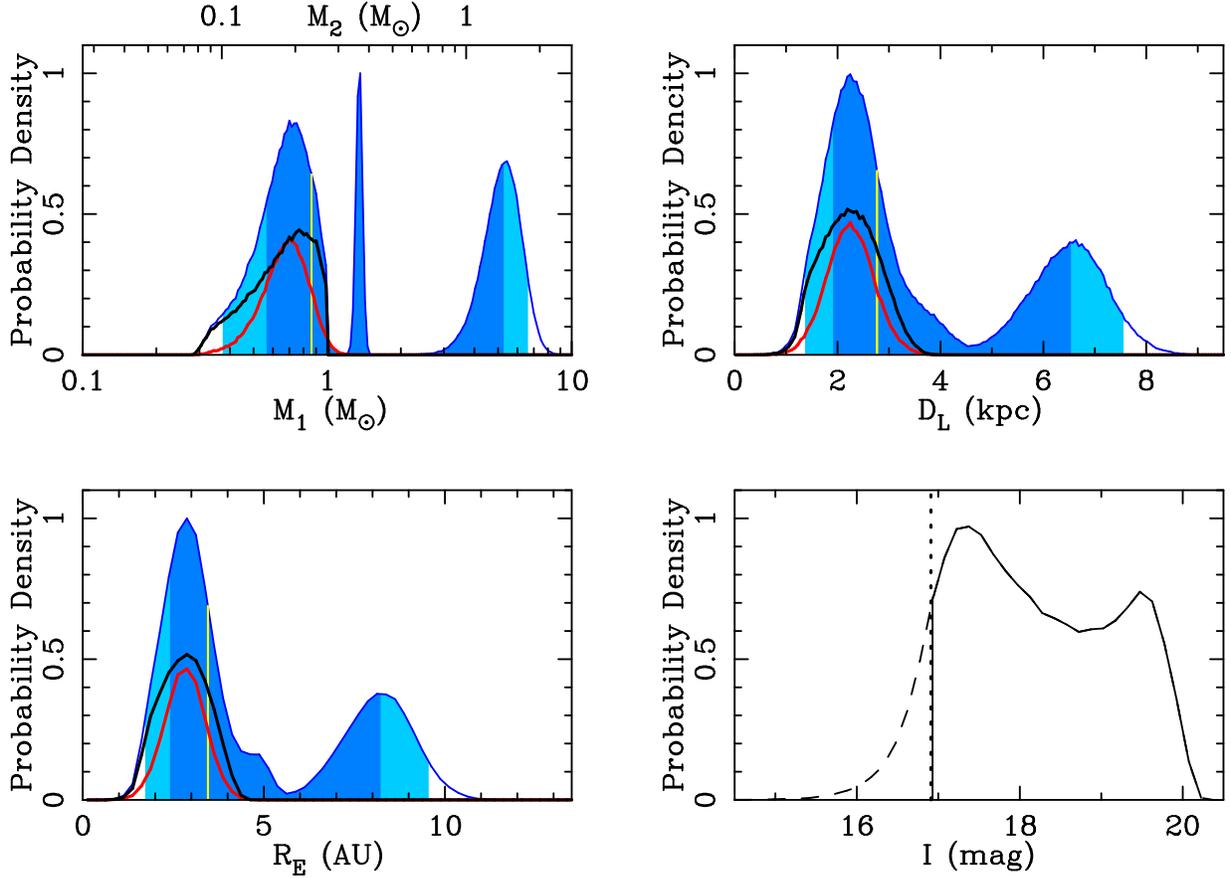} 
\caption{
The probability distribution from a Bayesian analysis with the upper limit of lens brightness 
for mass, $M_{\rm L}$, distance, $D_{\rm L}$, 
Einstein radius $R_{\rm E}$ and $I$-band magnitude of primary lens brightness. 
The vertical solid lines in yellow indicate the median values. 
The distribution described in blue indicates the results with the entire mass function 
and the curves in black and red represent the results for only main sequence stars and white dwarfs, respectively. 
The dark and light blue shaded regions indicate the 68\% and 95\% confidence limits, respectively. 
In the bottom-right panel, the upper limit of brightness, $I_{\rm b,0} > 16.91$ is shown by the vertical dotted line. 
The dashed curve indicates the rejected area due to the lens brightness constraint. 
Note that $D_{\rm S}$ is not fixed in the actual Bayesian analysis.}
\label{fig:likelihood}
\end{center}\end{figure}

The probability distribution from the Bayesian analysis is shown in Figure \ref{fig:likelihood}. 
The distribution described in blue indicates the results for all mass functions, 
and the curves in black and red represent the results for only from MSs and WDs, respectively. 
In the bottom-right panel, 
the probability distribution for the brightness is derived using only main sequence stars 
and excludes remnant stars as these should be very faint. 
The solid and dashed curve indicate the probability of the brightness with and without 
the upper limit of the brightness $I_{\rm b,0} > 16.91$, respectively, 
and the vertical dashed line represents the upper limit of the brightness.  
The probability ratio for each class of MSs, WDs, NSs, and BHs is 0.35, 0.23, 0.09, and 0.33, respectively, 
which indicates that the primary lens is likely a stellar remnant 65\% of the time, 
such as a white dwarf, a neutron star or a black hole. 

As a result, the lens properties were derived from the probability distribution as follows: 
The primary star is a massive star with a mass of $M_1 = 0.9_{-0.3}^{+4.6} M_\odot$, 
distance of $D_{\rm L} = 2.6_{-0.9}^{+3.8}$ kpc, and the companion has a mass of $M_2 = 0.2_{-0.1}^{+1.2} M_\odot$. 
The Einstein radius is $R_{\rm E} = 3.5_{-1.1}^{+5.2}$ AU, which means that 
the projected separation is $r_\perp = (0.3_{-0.1}^{+0.5}, \hspace{5pt} 56_{-18}^{+83})$ AU and 
the physical three-dimensional separation is $a = (0.4_{-0.2}^{+0.5}, \hspace{5pt} 82_{-37}^{+96})$ AU 
on the close and wide separation model, respectively. 
The physical three-dimensional separation was estimated by putting a planetary orbit 
at random inclination, eccentricity and phase \citep{GouldLoeb1992}. 

Recently, the mass functions of stellar remnants have been updated  
by improvements of observational methods and detections of stellar remnants 
(white dwarfs: \citet{Kepler2007}, neutron stars: \citet{Kiziltan2010}, and black holes: \citet{Ozel2010}).
However, the qualitative conclusions are not affected by these uncertain mass functions. 
Thus, we can conclude that the primary lens is most likely to be a stellar remnant 
such as a white dwarf, a neutron star or a black hole.

\section{Discussion and Conclusion}
\label{sec:conclusion}

We have reported the binary microlensing event OGLE-2007-BLG-514. 
This is an extremely high-magnification event ($A > 1000$), which enabled follow-up observations of this event. 
From the binary lens model search, we found that the lens has a stellar binary mass ratio $q \geq 0.1$. 
The well-known degeneracies of the close/wide separation and impact parameter $u_0>0$ and $u_0<0$ both remained unresolved 
in our light-curve analysis. 
We also search for the higher order effects of microlens parallax, lens orbital motion, 
and xallarap (source orbital motion) for this event. 
The xallarap model $\chi^2$ is better than the parallax and lens orbital motion model, respectively. 
However, spurious xallarap signals at this level are common in microlensing \citep{Poindexter2005}. 
And, the data on this event has high S/N, so we expect that stronger false signals can be caused by systematics. 
Therefore, the combined parallax and lens orbital motion model is the most plausible solution, 
but we can not absolutely exclude the xallarap model. 
Thus, we consider two solutions of the parallax model and xallarap model for the estimation of the lens properties. 
The best fit model with parallax and lens orbital motion has a binary mass ratio of $q = 0.321 \pm 0.007$ and 
a separation of $s = 0.072 \pm 0.001$ in a unit of Einstein radius with $\pi_{\rm E} = 0.13 \pm 0.02$.  
On the other hand, the xallarap and parallax model has a binary mass ratio of $q = 0.270 \pm 0.005$ 
and a separation of $s = 0.083 \pm 0.001$ Einstein radius. 

The parallax model parameters allow us to determine the lens properties. 
According to the model with parallax and lens orbital motion effects, 
the lens total mass is $M_{\rm L}=1.40 \pm 0.18 M_\odot$, as the components of the lens, the primary lens has 
$M_1= 1.06 \pm 0.13  M_\odot$ with a companion $M_2= 0.34 \pm 0.04 M_\odot$. 
The distance to the lens is $D_{\rm L} = 3.11 \pm 0.39~\mathrm{kpc}$. 
The upper limit of the brightness implies that the upper limit of the lens mass is $M < 0.99 M_\odot$, 
assuming the primary lens is a main sequence star. 
The main sequence star is the dominant component in the galaxy, 
and the upper limit of the lens mass, assuming that the primary lens is a main sequence star, 
is consistent within the error bars of the primary lens mass 
calculated by the parallax model parameters.
Thus, the parallax model indicates that the binary lens system is likely constructed by the main sequence stars. 

On the other hand, the probability distribution of the lens properties was estimated by the Bayesian analysis 
using the xallarap and parallax model. 
The parallax amplitude, $\pi_{\rm E} < 0.5$, is adopted for the upper limit of the lens mass. 
Also, the upper limit of the brightness in $I$ band was included in the analysis. 
As a results, the primary star is a massive star with a mass of $M_1 = 0.9_{-0.3}^{+4.6} M_\odot$, 
a distance of $D_{\rm L} = 2.6_{-0.9}^{+3.8}$ kpc, and a companion mass of $M_2 = 0.2_{-0.1}^{+1.2} M_\odot$.
The probability ratio for each class of MSs, WDs, NSs, and BHs is 0.35, 0.23, 0.09, 0.33, respectively, 
which indicates that the primary lens is likely a stellar remnant 65\% of the time, 
such as a white dwarf, a neutron star or a black hole. 

Moreover, based on the OGLE astrometry, we found that the blended light is associated with the event to high precision
(the difference of the position of the source and lens is estimated to be 23 mas, i.e., within the astrometric error.). 
Therefore, we considered three possibilities of where the blended light was coming from. 
The first is the primary lens. This suggestion is consistent with the parallax model that 
the primary lens is likely a main sequence star, which has a solar mass. 
The second is the companion of the lens. This is possible to explain by the results of the xallarap model.  
For example, if the primary lens is a black hole, the companion of the lens have a mass close to a solar mass. 
The last is the companion of the source in the xallarap model. 
But, it is difficult to interpret that the blended light is only coming from the companion of the source 
because the companion of the source is far from the Earth and very low mass, so the companion is very faint. 

As a final conclusion, we found two solutions for this event.
The one is the parallax model, which indicates that the binary lens system is likely constructed by the main sequence stars.
The another is the xallarap model, which indicates that the primary lens is likely the stellar remnant, 
such as a white dwarf, a neutron star or a black hole. 

In the future, follow-up observations by radio, optical, or X-ray space telescopes will identify 
whether the primary lens is a main sequence star or a stellar remnant. 
High spatial resolution observations with large ground-based telescope
and Adaptive-Optics instruments, such as the Subaru telescope or the
Very Large Telescope (VLT), could estimate more precise brightness from the lens
using the source brightness obtained from the modeling. 
In addition, it has been 4 years since the peak magnification of this event, 
and the lens-source relative proper motion is $2.70 \pm 0.15$ mas year$^{-1}$, indicating that 
the separation of the source and lens on the sky plane is about 10 mas by now. 
High resolution observation with the space telescope, Hubble Space Telescope (HST), 
could detect an elongated PSF blended with the lens and source, and verify the nature of the lens.

Microlensing searches are going on throughout the world. 
In particular, the OGLE group began using a new wide field of view camera, the OGLE-IV (1.4 deg$^2$ FOV), 
in March 2010 and the number of microlensing events dramatically increased after the OGLE-IV upgrade. 
Of course, the MOA group is continuing to operate its microlensing search with the MOA-II telescope 
with its wide field of view (2.2 deg$^2$) MOA-cam3 CCD camera \citep{Sako2008}. 
In 2011, two survey groups found about 1800 microlensing events, which constitutes a two fold increase over last year. 
More and more binary lens objects varying from planetary systems to massive binary systems 
can be discovered by further microlensing observation.

\acknowledgements
We acknowledge the following support: 
The OGLE project has received funding from the European Research Council
under the European Community's Seventh Framework Programme
(FP7/2007-2013) / ERC grant agreement no. 246678 to AU. 
The MOA project is supported by the Grant-in-Aid for Scientific Research 
(JSPS17340074, JSPS18253002), 
JSPS Research fellowships and the Global COE Program of Nagoya University 
``Quest for Fundamental Principles in the Universe" from JSPS and MEXT of Japan.
N.M. is supported by JSPS Research Fellowships for Young Scientists. 
RoboNet (KH,DB,MD,RAS,CS,YT) acknowledges support from
The Qatar Foundation via QNRF grant NPRP-09-476-1-78.

\clearpage

% TABLE

\begin{deluxetable}{lcc}
\tablecaption{Re-normalizing parameters for error bars \label{tb:norm}}
\tablewidth{0pt}
\tablehead{
\colhead{data set} & \colhead{$k$} & \colhead{$e_{\rm min}$} }
\startdata
OGLE                                    & 1.16 & 0.01 \\ % ogle
MOA-II                                  & 0.85 & 0.02 \\ % moa
$\mu$FUN Auckland                       & 0.93 & 0.00 \\ % auck
$\mu$FUN SMARTS CTIO $V$-band           & 3.00 & 0.00 \\ % ctiov 
$\mu$FUN SMARTS CTIO $I$-band           & 1.81 & 0.00 \\ % ctioi 
$\mu$FUN SMARTS CTIO $H$-band           & 1.00 & 0.01 \\ % ctioh 
$\mu$FUN Campo Catino Austral           & 0.60 & 0.00 \\ % cao 
$\mu$FUN Farm Cove                      & 0.92 & 0.00 \\ % farm
RoboNet Faulkes Telescope North         & 0.73 & 0.01 \\ % ftn
RoboNet Faulkes Telescope South         & 0.48 & 0.00 \\ % fts 
PLANET Canopus                          & 0.72 & 0.00 \\ % canopus 
IRSF $J$-band                           & 0.76 & 0.01 \\ % irsfj
IRSF $H$-band                           & 0.83 & 0.01 \\ % irsfh
IRSF $K_{\rm s}$-band                   & 0.76 & 0.01 \\ % ifsrk
\enddata
\tablecomments{The formula to re-normalize the error bars is $\sigma'_{i} = k \sqrt{\sigma_i^2 + e_{\rm min}^2}$, where $\sigma_i$ is original error bar of the $i$th data point within a unit of magnitude. }
\end{deluxetable}

\begin{deluxetable}{ccccccc}
\tablecaption{Limb darkening coefficients for the source star \label{tb:LDparameter}}
\tablewidth{0pt}
\tablehead{
\colhead{filter	color} & \colhead{$V$} & \colhead{$R$} & \colhead{$I$} & \colhead{$J$} & \colhead{$H$} & \colhead{$K$} }
\startdata
$u$             &   0.7242  &   0.6481  &   0.5618  &   0.4387  &   0.3711  &   0.3212 \\
\enddata
\tablecomments{These coefficients are for the source star with effective temperature $T_{\rm eff}$=5500~K, surface gravity $\log g$=4.0 cm s$^{-2}$ and metallicity $\log$[M/H]=0.3 \citep{Claret2000}. We used the $R$-band parameter for unfiltered data.}
\end{deluxetable}

\begin{deluxetable}{crrrrrrrrrr}
\tabletypesize{\tiny}
%\rotate
\tablecaption{Binary lens model parameters for parallax \label{tb:parallax}}
\tablewidth{0pt}
\tablehead{
\colhead{Model}           & \colhead{$t_0$}           & \colhead{$t_E$}            & \colhead{$u_0$}        & 
\colhead{$q$}             & \colhead{$s$}             & \colhead{$\alpha$}         & \colhead{$\rho$}       & 
\colhead{$\pi_{\rm E,N}$} & \colhead{$\pi_{\rm E,E}$} & \colhead{$\chi^2$} \\
\colhead{}                & \colhead{HJD'}            & \colhead{[days]}           & \colhead{$10^{-3}$}    & 
\colhead{}                & \colhead{}                & \colhead{[rad]}            & \colhead{$10^{-4}$}    & 
\colhead{}                & \colhead{}                & \colhead{} 
}
\startdata
close1.+ & 4386.784 & 198.96 & 1.60 & 0.249 & 0.0777 & 0.690 & 2.93 & 
          \nodata & \nodata & 3601.37 \\ 
         $\sigma$ & 0.001 & 3.14 & 0.02 & 0.007 & 0.0007 & 0.002 & 0.05 & 
          \nodata & \nodata & \\
         & & & & & & & & & & \\ 
close1.- & 4386.784 & 199.80 & -1.59 & 0.238 & 0.0787 & 5.595 & 2.91 & 
          \nodata & \nodata & 3601.50 \\ 
         $\sigma$ & 0.001 & 3.85 & 0.03 & 0.009 & 0.0008 & 0.002 & 0.06 & 
          \nodata & \nodata & \\
         & & & & & & & & & & \\ 
close2.+ & 4386.782 & 254.56 & 1.27 & 0.227 & 0.0709 & 0.679 & 2.29 & 
          -0.206 & 0.021 & 3580.37 \\ 
         $\sigma$ & 0.001 & 3.40 & 0.02 & 0.007 & 0.0009 & 0.002 & 0.05 & 
          0.013 & 0.006 & \\
         & & & & & & & & & & \\ 
close2.- & 4386.783 & 205.34 & -1.57 & 0.251 & 0.0765 & 5.603 & 2.85 & 
          0.221 & 0.042 & 3590.89 \\ 
         $\sigma$ & 0.001 & 4.81 & 0.03 & 0.008 & 0.0008 & 0.002 & 0.07 
          & 0.023 & 0.010 & \\
         & & & & & & & & & & \\ 
close3.+ & 4386.787 & 200.05 & 1.58 & 0.251 & 0.0773 & 0.690 & 2.95 & 
          -0.142 & -0.642 & 3593.49 \\ 
         $\sigma$ & 0.001 & 2.29 & 0.02 & 0.008 & 0.0009 & 0.003 & 0.03 & 
          0.203 & 0.159 & \\
         & & & & & & & & & & \\ 
close3.- & 4386.784 & 195.45 & -1.63 & 0.241 & 0.0790 & 5.599 & 2.99 & 
          -0.079 & -0.493 & 3598.00 \\ 
         $\sigma$ & 0.001 & 2.74 & 0.03 & 0.005 & 0.0007 & 0.003 & 0.04 & 
          0.232 & 0.250 & \\
         & & & & & & & & & & \\ 
close4.+ & 4386.782 & 246.01 & 1.31 & 0.219 & 0.0729 & 0.680 & 2.37 & 
          -0.194 & 0.018 & 3583.96 \\ 
         $\sigma$ & 0.001 & 3.58 & 0.02 & 0.009 & 0.0008 & 0.002 & 0.04 & 
          0.021 & 0.008 & \\
         & & & & & & & & & & \\ 
close4.- & 4386.784 & 198.00 & -1.62 & 0.262 & 0.0770 & 5.600 & 2.96 & 
          0.215 & 0.045 & 3590.66 \\ 
         $\sigma$ & 0.001 & 3.38 & 0.03 & 0.009 & 0.0007 & 0.002 & 0.05 & 
          0.026 & 0.010 & \\
         & & & & & & & & & & \\ 
\tableline 
& & & & & & & & & & \\ 
wide1.+ & 4386.782 & 232.96 & 1.36 & 0.421 & 17.3863 & 0.686 & 2.51 & 
          \nodata & \nodata & 3601.18 \\ 
         $\sigma$ & 0.001 & 3.58 & 0.02 & 0.016 & 0.2971 & 0.002 & 0.04 & 
          \nodata & \nodata & \\
         & & & & & & & & & & \\ 
wide1.- & 4386.782 & 235.81 & -1.35 & 0.450 & 17.8030 & 5.596 & 2.48 & 
          \nodata & \nodata & 3601.03 \\ 
         $\sigma$ & 0.001 & 4.04 & 0.02 & 0.026 & 0.3277 & 0.002 & 0.04 & 
          \nodata & \nodata & \\
         & & & & & & & & & & \\ 
wide2.+ & 4386.781 & 277.00 & 1.17 & 0.395 & 18.5607 & 0.677 & 2.11 & 
          -0.174 & 0.019 & 3578.93 \\ 
         $\sigma$ & 0.001 & 2.77 & 0.01 & 0.018 & 0.2442 & 0.002 & 0.02 & 
          0.012 & 0.008 & \\
         & & & & & & & & & & \\ 
wide2.- & 4386.781 & 238.43 & -1.35 & 0.479 & 18.1461 & 5.604 & 2.45 & 
          0.182 & 0.039 & 3590.21 \\ 
         $\sigma$ & 0.001 & 2.84 & 0.01 & 0.018 & 0.1547 & 0.003 & 0.02 & 
          0.022 & 0.006 & \\
         & & & & & & & & & & \\ 
wide3.+ & 4386.786 & 236.20 & 1.33 & 0.460 & 17.9249 & 0.688 & 2.49 & 
          -0.137 & -0.527 & 3592.68 \\ 
         $\sigma$ & 0.001 & 3.16 & 0.02 & 0.014 & 0.1940 & 0.003 & 0.03 & 
           0.135 & 0.120 & \\
         & & & & & & & & & & \\ 
wide3.- & 4386.782 & 237.32 & -1.34 & 0.402 & 17.3515 & 5.602 & 2.47 & 
          -0.079 & -0.400 & 3597.90 \\ 
         $\sigma$ & 0.001 & 4.11 & 0.02 & 0.013 & 0.1737 & 0.003 & 0.05 & 
          0.136 & 0.196 & \\
         & & & & & & & & & & \\ 
wide4.+ & 4386.781 & 264.15 & 1.22 & 0.418 & 18.4004 & 0.680 & 2.21 & 
          -0.170 & 0.020 & 3582.38 \\ 
         $\sigma$ & 0.001 & 3.06 & 0.01 & 0.011 & 0.1652 & 0.002 & 0.02 & 
          0.017 & 0.006 & \\
         & & & & & & & & & & \\ 
wide4.- & 4386.781 & 241.60 & -1.33 & 0.477 & 18.2562 & 5.605 & 2.43 & 
          0.180 & 0.035 & 3590.36 \\ 
         $\sigma$ & 0.001 & 3.32 & 0.02 & 0.014 & 0.1797 & 0.002 & 0.03 & 
          0.019 & 0.005 & \\
         & & & & & & & & & & \\ 
\enddata
\tablecomments{
Each model is classified by following characters. The character ``1" indicates a binary standard model, 
the characters ``2" and ``3" represent models with orbital or terrestrial parallax, respectively. 
The character ``4" indicates a model with both parallax effects. 
The names, ``close" and ``wide", indicate separations $s<1$ and $s>1$, respectively. 
For the $u_0>0$ model, we use the character ``+" and for the $u_0<0$ model we use the character ``-". 
The error bars represented as ``$\sigma$" are given by MCMC.  The $\chi^2$ value is the result of the fitting with 3588 data points. 
Note that the $u_0$ conventions are the same as in Figure 2 of \citep{Gould2004} and HJD$' \equiv$ HJD - 2,450,000.   
}
\end{deluxetable}

\begin{deluxetable}{crrrrrrrrrrrr}
\tabletypesize{\tiny}
\rotate
\tablecaption{Binary lens model parameters for lens orbital motion \label{tb:OM}}
\tablewidth{0pt}
\tablehead{
\colhead{Model}           & \colhead{$t_0$}           & \colhead{$t_E$}            & \colhead{$u_0$}        & 
\colhead{$q$}             & \colhead{$s$}             & \colhead{$\alpha$}         & \colhead{$\rho$}       & 
\colhead{$\pi_{\rm E,N}$} & \colhead{$\pi_{\rm E,E}$} & \colhead{$\omega$}         & \colhead{$ds/dt$}      & \colhead{$\chi^2$} \\
\colhead{}                & \colhead{HJD'}            & \colhead{[days]}           & \colhead{$10^{-3}$}    & 
\colhead{}                & \colhead{}                & \colhead{[rad]}            & \colhead{$10^{-4}$}    & 
\colhead{}                & \colhead{}                & \colhead{10$^{-2}$}        & \colhead{10$^{-2}$}    & \colhead{} \\
\colhead{}                & \colhead{}                & \colhead{}                 & \colhead{}             & 
\colhead{}                & \colhead{}                & \colhead{}                 & \colhead{}             & 
\colhead{}                & \colhead{}                & \colhead{[rad days$^{-1}$]}& \colhead{[days$^{-1}$]}& \colhead{} 
}
\startdata
  & & & & & & & & & & & & \\ 
  \multicolumn{13}{l}{lens orbital motion} \\
  \tableline 
          & & & & & & & & & & & & \\ 
close.+ & 4386.784 & 170.62 & 1.94 & 0.376 & 0.076 & 0.647 & 3.53 &
          \nodata & \nodata & -4.9390 & 1.00 & 3553.26 \\
          $\sigma$ & 0.001 & 1.60 & 0.02 & 0.014 & 0.001 & 0.003 & 0.03 & 
          \nodata & \nodata & 0.1753 & 0.20 &  \\
          & & & & & & & & & & & & \\ 
close.- & 4386.785 & 178.76 & -1.85 & 0.353 & 0.075 & 5.633 & 3.37 &
          \nodata & \nodata & 4.9103 & 0.87 & 3551.59 \\
          $\sigma$ & 0.001 & 1.91 & 0.02 & 0.008 & 0.001 & 0.003 & 0.04 & 
          \nodata & \nodata & 0.1955 & 0.15 &  \\
          & & & & & & & & & & & & \\ 
wide.+ & 4386.746 & 225.95 & 1.38 & 0.427 & 17.552 & 0.674 & 2.48 &
          \nodata & \nodata & -0.0015 & -0.57 & 3601.19 \\
          $\sigma$ & 0.009 & 3.72 & 0.03 & 0.024 & 0.246 & 0.010 & 0.03 & 
          \nodata & \nodata & 0.0006 & 0.16 &  \\
          & & & & & & & & & & & & \\ 
wide.- & 4386.712 & 215.09 & -1.24 & 0.447 & 17.874 & 5.570 & 2.45 &
          \nodata & \nodata & 0.0070 & -0.89 & 3600.81 \\
          $\sigma$ & 0.006 & 2.67 & 0.02 & 0.015 & 0.152 & 0.009 & 0.02 & 
          \nodata & \nodata & 0.0006 & 0.10 &  \\
          & & & & & & & & & & & & \\ 
  \tableline 
  & & & & & & & & & & & & \\ 
  \multicolumn{13}{l}{parallax + lens orbital motion} \\
  \tableline 
          & & & & & & & & & & & & \\ 
close.+ & 4386.781 & 250.29 & 1.32 & 0.311 & 0.066 & 0.651 & 2.38 &
          -0.141 & -0.010 & -3.5485 & 1.02 & 3549.82 \\
          $\sigma$ & 0.001 & 3.42 & 0.03 & 0.012 & 0.001 & 0.002 & 0.06 & 
          0.012 & 0.003 & 0.0051 & 0.02 & \\
          & & & & & & & & & & & & \\ 
close.- & 4386.781 & 202.40 & -1.64 & 0.321 & 0.072 & 5.644 & 2.98 &
          0.131 & 0.004 & 4.8568 & 0.94 & 3548.29 \\
          $\sigma$ & 0.001 & 2.28 & 0.02 & 0.007 & 0.001 & 0.002 & 0.04 &  
          0.015 & 0.002 & 0.0035 & 0.01 & \\
          & & & & & & & & & & & & \\ 
wide.+ & 4386.872 & 292.38 & 0.96 & 0.399 & 18.358 & 0.759 & 2.17 &
          -0.164 & 0.032 & -0.0004 & 1.42 & 3583.05 \\
          $\sigma$ & 0.007 & 3.14 & 0.02 & 0.019 & 0.207 & 0.006 & 0.03 & 
          0.015 & 0.007 & 0.0001 & 0.08 & \\
          & & & & & & & & & & & & \\ 
wide.- & 4386.943 & 288.83 & -1.26 & 0.447 & 17.936 & 5.582 & 2.42 &
          0.180 & 0.040 & -0.0066 & 1.85 & 3590.67 \\
          $\sigma$ & 0.004 & 3.64 & 0.02 & 0.019 & 0.185 & 0.007 & 0.03 & 
          0.016 & 0.005 & 0.0004 & 0.05 & \\
          & & & & & & & & & & & & \\ 
\enddata
\tablecomments{
The names, ``close" and ``wide", indicate separations $s<1$ and $s>1$, respectively. 
For the $u_0>0$ model, we use the character ``+" and for the $u_0<0$ model we use the character ``-". 
The error bars represented as ``$\sigma$" are given by MCMC.  The $\chi^2$ value is the result of the fitting with 3588 data points. 
Note that the $u_0$ conventions are the same as in Figure 2 of \citep{Gould2004} and HJD$' \equiv$ HJD - 2,450,000.   
}
\end{deluxetable}

\begin{deluxetable}{crrrrrrrrrrrrrrr}
\tabletypesize{\tiny}
\rotate
\tablecaption{Binary lens model parameters for xallarap \label{tb:xallarap}}
\tablewidth{0pt}
\tablehead{
\colhead{Model}           & \colhead{$t_0$}           & \colhead{$t_E$}            & \colhead{$u_0$}        & 
\colhead{$q$}             & \colhead{$s$}             & \colhead{$\alpha$}         & \colhead{$\rho$}       & 
\colhead{$\pi_{\rm E,N}$} & \colhead{$\pi_{\rm E,E}$} & \colhead{$\xi_{\rm E,N}$}  & \colhead{$\xi_{\rm E,E}$}  & 
\colhead{RA$_\xi$}        & \colhead{decl$_\xi$}      & \colhead{P$_\xi$}          & \colhead{$\chi^2$} \\
\colhead{}                & \colhead{HJD'}            & \colhead{[days]}           & \colhead{$10^{-3}$}    & 
\colhead{}                & \colhead{}                & \colhead{[rad]}            & \colhead{$10^{-4}$}    & 
\colhead{}                & \colhead{}                & \colhead{$10^{-4}$}        & \colhead{$10^{-4}$}    & 
\colhead{[deg]}           & \colhead{[deg]}           & \colhead{[days]}           & \colhead{} 
}
\startdata

  & & & & & & & & & & & & \\ 
  \multicolumn{16}{l}{xallarap} \\
  \tableline 
          & & & & & & & & & & & & \\ 
close.+ & 4386.789 & 153.18 & 2.12 & 0.283 & 0.0860 & 0.683 & 3.83 &
          \nodata & \nodata & -1.27 & 4.32 & 210.6 & 13.95 & 9.992 & 3546.78 \\
          $\sigma$ & 0.001 & 1.09 & 0.01 & 0.006 & 0.0005 & 0.004 & 0.02 & 
          \nodata & \nodata & 0.09 & 0.11 & 10.1 & 11.28 & 0.007 &  \\
          & & & & & & & & & & & & \\ 
close.- & 4386.784 & 157.59 & -2.11 & 0.291 & 0.0842 & 5.616 & 3.73 &
          \nodata & \nodata & 4.10 & 3.17 & 250.4 & -63.43 & 9.992 & 3547.27 \\
          $\sigma$ & 0.001 & 1.08 & 0.02 & 0.006 & 0.0005 & 0.003 & 0.02 & 
          \nodata & \nodata & 0.15 & 0.18 & 13.3 & 9.08 & 0.011 & \\
          & & & & & & & & & & & & \\ 
wide.+ & 4386.786 & 187.96 & 1.74 & 0.530 & 16.4576 & 0.675 & 3.12 &
          \nodata & \nodata & -3.67 & 2.55 & 203.0 & -6.14 & 9.737 & 3546.42 \\
          $\sigma$ & 0.001 & 1.24 & 0.01 & 0.014 & 0.1173 & 0.003 & 0.01 & 
          \nodata & \nodata & 0.32 & 0.09 & 14.2 & 6.12 & 0.009 & \\
          & & & & & & & & & & & & \\ 
wide.- & 4386.786 & 185.51 & -1.75 & 0.493 & 16.0743 & 5.608 & 3.17 &
          \nodata & \nodata & 2.91 & 2.74 & 16.4 & 1.14 & 9.366 & 3544.88 \\
          $\sigma$ & 0.001 & 1.04 & 0.01 & 0.012 & 0.0994 & 0.003 & 0.02 & 
          \nodata & \nodata & 0.21 & 0.13 & 10.3 & 7.27 & 0.006 & \\
          & & & & & & & & & & & & \\ 
  \tableline 
  & & & & & & & & & & & & \\ 
  \multicolumn{16}{l}{xallarap + parallax} \\
  \tableline 
          & & & & & & & & & & & & \\ 
close.+ & 4386.788 & 169.39 & 1.93 & 0.270 & 0.0827 & 0.677 & 3.45 &
          0.212 & -0.091 & -2.66 & 2.86 & 247.6 & -7.36 & 9.139 & 3535.83 \\
          $\sigma$ & 0.001 & 1.02 & 0.01 & 0.005 & 0.0007 & 0.003 & 0.02 & 
          0.025 & 0.015 & 0.11 & 0.14 & 14.1 & 7.84 & 0.006 & \\
          & & & & & & & & & & & & \\ 
close.- & 4386.788 & 162.58 & -2.04 & 0.264 & 0.0850 & 5.608 & 3.61 &
          0.213 & -0.064 & 6.75 & 3.93 & 447.1 & -163.07 & 9.158 & 3536.47 \\
          $\sigma$ & 0.001 & 1.36 & 0.01 & 0.006 & 0.0007 & 0.004 & 0.03 & 
          0.032 & 0.023 & 0.15 & 0.20 & 15.2 & 9.75 & 0.008 & \\
          & & & & & & & & & & & & \\ 
wide.+ & 4386.785 & 189.06 & 1.73 & 0.519 & 16.4318 & 0.670 & 3.10 &
          0.200 & -0.096 & -1.58 & 4.41 & 344.4 & 5.79 & 9.347 & 3537.02 \\
          $\sigma$ & 0.001 & 1.27 & 0.01 & 0.011 & 0.0779 & 0.003 & 0.02 & 
          0.024 & 0.018 & 0.11 & 0.15 & 3.4 & 5.90 & 0.003 & \\
          & & & & & & & & & & & & \\ 
wide.- & 4386.786 & 193.58 & -1.68 & 0.454 & 16.0947 & 5.605 & 3.04 &
          0.195 & -0.084 & 1.67 & 4.34 & 173.2 & 15.35 & 9.979 & 3539.78 \\
          $\sigma$ & 0.001 & 1.74 & 0.02 & 0.010 & 0.0946 & 0.003 & 0.03 & 
          0.031 & 0.018 & 0.16 & 0.10 & 5.4 & 9.18 & 0.005 \\
          & & & & & & & & & & & & \\ 
\enddata
\tablecomments{
The names, ``close" and ``wide", indicate separations $s<1$ and $s>1$, respectively.  
For the $u_0>0$ model, we use the character ``+" and for the $u_0<0$ model we use the character ``-". 
The error bars represented as ``$\sigma$" are given by MCMC.  The $\chi^2$ value is the result of the fitting with 3588 data points. 
Note that the $u_0$ conventions are the same as in Figure 2 of \citep{Gould2004} and HJD$' \equiv$ HJD - 2,450,000. 
}
\end{deluxetable}

\begin{deluxetable}{lrrrrrrrr}
\tabletypesize{\footnotesize}
%\rotate
\tablecaption{Source and Blending brightness \label{tb:flux}}
\tablewidth{0pt}
\tablehead{
\colhead{data}      & \colhead{$F_{\rm s}$} & \colhead{$F_{\rm s, err}$} & \colhead{$F_{\rm b}$} & \colhead{$F_{\rm b, err}$} & 
\colhead{Source}    & \colhead{error} & \colhead{Blending}     & \colhead{error} \\
\colhead{}          & \colhead{[ADU]}      & \colhead{}             & \colhead{[ADU]}      & \colhead{}             & 
\colhead{Magnitude} & \colhead{}      & \colhead{Magnitude}    & \colhead{}       
}
\startdata
OGLE $I$  &  12.63 & 0.10 &  62.4 &   3.0 & 21.25 & 0.01 &   19.51 &    0.04  \\
CTIO $V$  &   0.83 & 0.02 & -16.9 &   8.7 & 24.20 & 0.03 & \nodata & \nodata  \\
CTIO $I$  &  10.58 & 0.13 &  50.3 &  27.9 & 21.44 & 0.01 &   19.75 &    0.60  \\
CTIO $H$  & 103.16 & 1.95 &  2012 &   402 & 18.97 & 0.02 &   15.74 &    0.22  \\
IRSF $J$  &  58.37 & 0.30 &   103 &   128 & 19.59 & 0.01 &   18.97 &    1.35  \\
IRSF $H$  & 113.28 & 0.51 &   382 &   211 & 18.87 & 0.01 &   17.54 &    0.60  \\
IRSF $K_{\rm s}$ & 146.48 & 0.69 & 1301 & 294 & 18.59 & 0.01 &  16.21 &  0.25  \\
\enddata
\tablecomments{The photometry file for each data set excluding the OGLE data was made by the DoPHOT tool 
to estimate blending fluxes and source fluxes. 
}
\end{deluxetable}

\begin{deluxetable}{lrrrrrr}
\tabletypesize{\footnotesize}
%\rotate
\tablecaption{The upper limit of the lens brightness \label{tb:upperlimit}}
\tablewidth{0pt}
\tablehead{
\colhead{data} & \colhead{$f_{\rm sky}$} & \colhead{$r$} & \colhead{$G$} & \colhead{$c$} & \colhead{M$_{\rm limit}$} & \colhead{M$_{\rm limit,blend}$} \\
\colhead{}   & \colhead{[ADU]} & \colhead{[pixel]} & \colhead{[$e^{-}$/ADU]} & \colhead{} & \colhead{} & \colhead{} 
}
\startdata
OGLE $I$         &  \nodata & \nodata  & \nodata & \nodata & \nodata & 19.47 \\
CTIO $V$         &   379.2 & 5 & 2.3 & 27.87$\pm$0.08 & 20.90 & \nodata \\
CTIO $I$         &   609.4 & 5 & 2.3 & 26.91$\pm$0.18 & 19.59 & 19.15 \\
IRSF $J$         &  1297.7 & 4 & 5.0 & 24.32$\pm$0.09 & 17.33 & 17.62 \\
IRSF $H$         &  5427.7 & 4 & 5.0 & 24.35$\pm$0.08 & 16.61 & 16.96 \\
IRSF $K_{\rm s}$ &  5955.6 & 3 & 5.0 & 23.50$\pm$0.08 & 15.97 & 15.96 \\\enddata
\tablecomments{The $f_{\rm sky}$ is sky flux at this target position, 
$r$ is the radius of a PSF area, and $c$ is the scale factor to calibrate to catalog magnitude. 
M$_{\rm limit}$ and M$_{\rm limit, blend}$ indicate the upper limit of the lens brightness 
estimated by the flux in the images observed in 2011 and the blending magnitude obtained from the fit, respectively.}
\end{deluxetable}

\end{document}